\documentstyle[prl,aps]{revtex}
\draft
\newcommand{\med}[1]{\langle #1 \rangle}
\newtheorem{theorem}{Theorem}
\begin{document}

\twocolumn[\hsize\textwidth\columnwidth\hsize\csname@twocolumnfalse\endcsname

\title{The Kac limit for finite-range spin glasses}

\author{Silvio Franz~\cite{sf},
Fabio Lucio Toninelli~\cite{ft}}

\address{ \cite{sf} The Abdus Salam
International Center for Theoretical Physics, Condensed Matter Group\\
Strada Costiera 11, P.O. Box 586, I-34100 Trieste, Italy\\ 
\cite{ft} EURANDOM, Technische Universiteit Eindhoven, P.O. Box
513, 5600 MB Eindhoven, The Netherlands\\}

\date{\today}

\maketitle

\begin{abstract}
 We consider a finite range spin glass model in arbitrary dimension,
where the strength of the two-body coupling decays to zero 
over some distance $\gamma^{-1}$.  We show that, under mild
assumptions on the interaction potential, the infinite-volume 
free energy of the system converges to that of the Sherrington-Kirkpatrick
one, in the Kac limit $\gamma\to0$.
This could be a first step toward an expansion around mean field theory, for spin glass systems.
\end{abstract}


\twocolumn\vskip.5pc]\narrowtext


Despite years of debate, the nature of the spin glass phase of the
finite dimensional systems remains a major open problem in statistical
physics. Two competing theories have been proposed as candidate to
explain spin glass physics at low temperature: the theory of replica
symmetry breaking \cite{MPV} \cite{petc} and the droplet theory
\cite{fh} \cite{BrMo}. The former, based on the analysis of the long
range Sherrington-Kirkpatrick (SK) spin glass, predicts a rich
phenomenology with ergodicity breaking not related to any physical
symmetry breaking and susceptibility anomalies related to the presence
on many pure states. The latter assimilates spin glasses to some kind
of "disguised ferromagnet" -albeit with complex phenomenology- where
the transition appears as a conventional symmetry breaking phenomenon.
Both theories being non-rigorous in the applications to finite
dimensional systems, it appears very difficult to solve the question
on a purely theoretical ground. On the other hand, experiments in 3D
and numerical simulations in 3 and 4D fail to give compelling evidence
in favour of one or the other of the two theories: the times probed in
the experiments are too short to settle the question of the presence
or absence of replica symmetry breaking and the related issue of
asymptotic existence of response anomalies during aging dynamics, and
the length scales probed in the simulations are too small to infer the
behaviour of the thermodynamic limit.  Rigorous analysis of finite
dimensional systems turns out to be very hard, and so far has not been
able to exclude either scenario, although it has produced \cite{ns}
considerable conceptual clarification, and shown some of the
subtleties hidden even in the definition of the infinite volume limit
of these models.  Even at the mean field level, only very recently,
simple interpolation methods have been introduced \cite{limterm}
\cite{broken} \cite{ass} which have allowed to prove \cite{talaparisi}
the Parisi solution for the SK model. Interpolation methods have
subsequently been applied also in the context of finite range spin
glasses, {\em e.g.} in \cite{cg}.

In this Letter we focus our attention on the Kac limit of finite range
spin glasses as first considered in \cite{froelich}, and later studied
in \cite{bovier-kac} and \cite{kacnoi}. Kac models are a classical
tool of mathematical physics, where one considers variables
interacting via a potential with finite range $\xi=\gamma^{-1}$, which
tends to infinity {\em after} the thermodynamic limit is taken. In a
classical paper \cite{lp} Penrose and Lebowitz proved that for
conventional non-disordered systems, the free-energy tends (modulo the
Maxwell construction) to the one of the corresponding mean-field
system where the interactions do not decay with distance and scale
with the size of the system.  We combine here the idea of the
interpolating model with the idea \cite{lp} of dividing the system
into boxes of suitable size to prove the same property in spin
glasses. 

Other disordered models with Kac-type interactions have been studied 
in previous literature. For instance, see \cite{bgp} and references therein
for the case of the Hopfield model.


The model we consider is defined on the $d$-dimensional
lattice ${ Z}^d$,  with 
Ising spin degrees of freedom $\sigma_i=\pm 1, i\in{ Z}^d$. 
Given a finite hypercube $\Lambda$ of side $L$ 
one defines the finite volume Hamiltonian as
\begin{eqnarray}
\label{hkac}
H^{(\gamma)}_{\Lambda}(\sigma,h;J)=-
\sum_{i,j\in\Lambda}
\sqrt{\frac{{w(i-j;\gamma)}}{{2 W(\gamma)}}}
J_{ij}\sigma_i\sigma_j-h\sum_{i\in \Lambda}\sigma_i,
\end{eqnarray}
where
$W(\gamma)=\sum_{i\in { Z}^d}w(i;\gamma)$
and
$w(r;\gamma)=\gamma^d \phi(\gamma r)$
for some smooth, nonnegative function $\phi(r)$, decaying sufficiently
fast for $|r|\to\infty$ to have $W(\gamma)<\infty$. The parameter
$\gamma=\xi^{-1}$ is the inverse range of the interaction. The
quenched couplings $J_{ij}$ are i.i.d.  Gaussian $N(0,1)$ variables,
and we denote by $E$ the corresponding averages.  As is well known
\cite{vuiller} \cite{vanenter}, the infinite-volume limit of the
quenched free energy
\begin{equation}
f^{(\gamma)}(\beta,h)=-\lim_{L\to\infty}
\frac1{\beta|\Lambda|}E\ln Z^{(\gamma)}_\Lambda(\beta,h;J)
\end{equation}
exists. 

On the other hand, the Hamiltonian of the 
SK
spin glass mean field model is defined as \cite{sk}
\begin{eqnarray}
\label{hsk}
H^{S.K.}_{|\Lambda|}(\sigma,h;J)=-\frac1{\sqrt{2|\Lambda|}}
\sum_{i,j\in \Lambda} J_{ij}
\sigma_i\sigma_j-h\sum_{i\in\Lambda}\sigma_i,
\end{eqnarray}
where $|\Lambda|=L^d$ is the number of lattice sites in $\Lambda$.
Subadditivity of the corresponding free energy 
and existence of its infinite volume limit
\begin{eqnarray}
f^{S.K.}(\beta,h)=-\lim_{L\to\infty}
\frac1{\beta|\Lambda|}E\ln Z^{S.K.}_{|\Lambda|}(\beta,h;J)
\end{eqnarray}
has been proven in \cite{limterm}.

It was recently shown in \cite{kacnoi} that the free energy of  model
(\ref{hkac}) 
is bounded below by that of SK:
\begin{equation}
\label{risGT}
f^{(\gamma)}(\beta,h)\ge f^{S.K.}(\beta,h)
\end{equation}
for any value of $d,\beta,h$ and $\gamma$, provided that 
the potential $\phi(i-j)$ is {\it nonnegative definite}, {\em i.e.},
its Fourier transform is nonnegative. 
For instance, it is immediate to check this condition for
$
\phi(i-j)=e^{-\sum_{\alpha=1}^d|i_\alpha-j_\alpha|},
$
which for $d=1$ is just
the potential considered originally by Kac in \cite{kac}.
In the present paper, we provide the complementary bound, which allows to 
fully characterize the quenched free energy in the Kac limit $\gamma\to0$:
\begin{theorem}
 Assume that $\sum_{i\in Z^d} \phi(i)<\infty$. Then, for any
 $\beta$ and $h$ one has
 \begin{eqnarray}
   \lim_{\gamma\to0} f^{(\gamma)}(\beta,h)\le f^{S.K.}(\beta,h).
 \end{eqnarray}
If in addition all the Fourier components of $\phi$ are  nonnegative, then
\begin{eqnarray}
   \lim_{\gamma\to0} f^{(\gamma)}(\beta,h)= f^{S.K.}(\beta,h).
 \end{eqnarray}
\end{theorem}
Together with Talagrand's recently established proof \cite{talaparisi}
of the Parisi ansatz for the SK model,
this shows that the Parisi theory \cite{MPV} gives the correct free energy 
for finite dimensional spin glasses in the Kac limit.

The idea of the proof 
is to interpolate between the Kac model in a volume $|\Lambda|$ and
a system made of a collection of many independent SK subsystems of
volume $M=\ell^d$. The crucial point, as in \cite{lp}, is to choose
\begin{equation}
\label{scale}
\ell \ll \xi \ll L,
\end{equation}
and to let  the  three lengths diverge in this order.
Let us divide the box $\Lambda$ into
sub-cubes $\Omega_n$ of volume $M$, $n=1,\cdots,|\Lambda|/M$, and 
introduce the interpolating partition function
\begin{eqnarray}
\nonumber
Z_\Lambda(t)&=&\sum_{\sigma}\exp\left(\beta\sqrt{1-t}\sum_n\sum_{i,j\in\Omega_n}
\frac{J_{ij}}{\sqrt{2M}}\sigma_i\sigma_j
\right)\\
\nonumber
& &\times\exp\left(\beta\sqrt t\sum_{i,j\in\Lambda}
\sqrt{\frac{w(i-j;\gamma)}{2W(\gamma)}}J'_{ij}\sigma_i\sigma_j
+\beta h\sum_{i\in\Lambda}\sigma_i\right),
\end{eqnarray}
where the Gaussian variables $J'$ are independent of 
the $J$.
Note that
\begin{eqnarray}
&& \frac1{|\Lambda|}E \ln Z_\Lambda(0)=\frac1M E\ln Z_M^{S.K.}(\beta,h;J)\\
&&\frac1{|\Lambda|}E \ln Z_\Lambda(1)=\frac1{|\Lambda|}
E\ln Z_\Lambda^{(\gamma)}(\beta,h;J).
\end{eqnarray}
As we show below, one has
\begin{eqnarray}
\label{bellissima}
\lim_{\gamma\to0}\lim_{L\to \infty}
\frac d{dt} \frac1{|\Lambda|}E\ln Z_\Lambda(t)\ge0
\end{eqnarray}
uniformly for $0\le t \le1$.  After integration on $t$ between $0$ and
$1$ and taking the large $M$ limit, one finds therefore the desired
result
\begin{eqnarray}
-\beta\lim_{\gamma\to0}f^{(\gamma)}(\beta,h)& \ge &
\lim_{M\to\infty}\frac1M E\ln Z_M^{S.K.}(\beta,h;J)\\
\nonumber 
& =&-\beta f^{S.K.}(\beta,h).
\end{eqnarray}
Denoting as  $\med.$ the Gibbs average,
the computation of the $t$ derivative gives, up to terms
negligible for large $L$,
\begin{eqnarray}
\frac d{dt} \frac1{|\Lambda|}E\ln Z_\Lambda(t)&= &\frac{\beta^2}{4|\Lambda|}
E\left[\sum_n\sum_{i,j\in \Omega_n}\frac1{M}\med{\sigma_i\sigma_j}^2
\right.
\label{deriv}
\\
\nonumber
& &
\left.
-\sum_{i,j\in\Lambda}\frac{w(i-j;\gamma)}{W(\gamma)}
\med{\sigma_i\sigma_j}^2
\right],
\end{eqnarray}
where we have used integration by parts on the Gaussian disorder 
and the property
\begin{equation}
\lim_{L\to \infty}\frac1{|\Lambda|}\sum_{i,j\in\Lambda}
\frac{w(i-j;\gamma)}{W(\gamma)}=1.
\end{equation}
Introducing two replicas with identical quenched
couplings and spin configurations $\sigma^1,\sigma^2$, 
we can write (\ref{deriv}) as:
\begin{eqnarray}
\label{form1}
\frac d{dt} \frac1{|\Lambda|}E\ln Z_\Lambda(t)= & &\frac{\beta^2}{4|\Lambda|}
E\left[
\sum_{n}\frac1M\sum_{i,j\in\Omega_n}
\med{\sigma^1_i\sigma^2_i\sigma^1_j\sigma^2_j}
\right. 
\\\nonumber
& & 
\left. 
-\sum_{i,j\in\Lambda}
\frac{w(i-j;\gamma)}{W(\gamma)}
\med{\sigma^1_i\sigma^2_i\sigma^1_j\sigma^2_j}
\right]. 
\end{eqnarray}
Denoting the partial overlap in the $n$-th sub-cube as
$
q_{12}^{(n)}=1/M\sum_{i\in \Omega_n}\sigma^1_i\sigma^2_i,
$
the first term of the r.h.s. can be rewritten as
\begin{equation}
\label{form2}
\frac{\beta^2M}{4|\Lambda|}\sum_nE\med{(q_{12}^{(n)})^2}.
\end{equation}
As for the second term, defining
\begin{equation}
w^+_{mn}=\sup_{i\in\Omega_m,j\in\Omega_n}\frac{w(i-j;\gamma)}{W(\gamma)}
\end{equation}
and using the straightforward inequality $2x y\le x^2+y^2$, one has
\begin{eqnarray}
\label{aux}
& &  \frac1{|\Lambda|}\sum_{i,j\in\Lambda}\frac{w(i-j;\gamma)}{W(\gamma)}
E\med{\sigma^1_i\sigma^2_i\sigma^1_j\sigma^2_j}
\\
\nonumber
& &
\le \frac{M^2}{2|\Lambda|}\sum_{n,m}w^+_{mn}E\med{(q_{12}^{(n)})^2
+(q_{12}^{(m)})^2}.
\end{eqnarray}
In the Kac limit $\gamma\to0$, the diagonal terms $n=m$ give a vanishing 
contribution. As for the nondiagonal ones, one observes 
that 
\begin{equation}
\lim_{\gamma\to0}\sum_{m(\ne n)} w^+_{mn}=\frac1M,
\end{equation}
where the summation runs only on one of the two indices, 
so that finally the r.h.s. of 
(\ref{aux}) is bounded above by
\begin{eqnarray}
  \frac M{|\Lambda|}\sum_nE\med{(q_{12}^{(n)})^2},
\end{eqnarray}
apart from a negligible error term.
Together with Eqs. (\ref{form1}) and (\ref{form2}), this proves 
(\ref{bellissima}) and therefore the Theorem.

As a side remark, it is easy to employ this  method, 
together with that of \cite{kacnoi}, to obtain a new  proof
of the existence of the thermodynamic limit for the SK model, independent
of the convexity  argument developed in \cite{limterm}. 

It is possible to generalize this theorem to the ``diluted Kac spin glass'' 
case \cite{kacnoi} where each given spin $\sigma_i$
interacts with a finite random number of 
other spins $\sigma_j$, which are chosen randomly according to a
probability distribution that decays to zero on the scale $\xi$,
as $|i-j|$ diverges.
In the Kac limit $\xi\to
\infty$, one can prove that the free energy of the model converges to 
that of its mean field counterpart, which in that case is the
Viana-Bray model \cite{viana}. 
Full details of the proof are given in \cite{kacvb}.

A second generalization of our result is to consider two replicas of
the system, coupled via a term depending on their mutual overlap. This
problem has been considered for instance in \cite{franzparisi} and is
relevant for the study of glassy dynamics, especially if applied to
models which exhibit ``one-step replica symmetry breaking''
\cite{MPV}.  The new feature here is that, at the mean field level,
the free energy of the coupled system can be expressed
\cite{franzparisi} in terms of an effective potential depending on the
overlap, which turns out to be {\em nonconvex}. It was argued in
\cite{fp} that a minimal modification of the theory in finite
dimension requires restoration of the convexity through the
Maxwell construction. This, analogously to the ordered case \cite{lp},
emerges naturally in the Kac limit of finite range models. We plan to
report on this soon \cite{aes}.

The main interest of the result presented in this Letter is that it
could represent for spin glasses, a first step toward an expansion
around the mean field case, which would hopefully shed some light on
the nature of the spin glass phase for models with finite -albeit
large- interaction range.  This hope is supported by 
the fact that a similar 
program has been successfully carried on recently
for non-random ferromagnetic spin systems \cite{cp} \cite{bz} \cite{bp}
and continuous particle systems \cite{lpres1},
showing that in dimension $d\ge2$ it is
possible to write a controlled  expansion around the 
$\gamma=0$ point, and to prove rigorously that for large but finite $\xi$ the system has a phase transition (broken spin flip or liquid-vapor, 
respectively) with coexisting phases.

{\bf  Acknowledgments}

We would like to thank Francesco Guerra for many enlightening
conversations.  F.L.T. is grateful to the Condensed Matter Group of the
ICTP for kind hospitality during the preparation of this work.  This
work was supported in part by the European Community's Human Potential
programme under contract ``HPRN-CT-2002-00319 STIPCO''.

\end{document}